\documentclass{elsart}
\usepackage{amssymb}
\usepackage{graphicx}

\def\lbullet{\mbox{\large $\bullet$}}

\def\ldiamond{\mbox{\large $\diamond$}}

\def\3{\ss }           

\def\bsigma{\mbox{\boldmath $\sigma$}}

\begin{document}

\begin{frontmatter}


\title{Constructing thermodynamically consistent models with a non-ideal 
equation of state}

\author[physicsuofm,supercomp]{Erkan T{\"u}zel}
\author[physicsndsu]{Thomas Ihle}
\author[supercomp,physicsndsu]{Daniel M. Kroll}

\address[physicsuofm]{School of Physics and Astronomy, University of Minnesota, 116 Church Street SE, , Minneapolis, MN 55455, USA.}
\address[supercomp]{Supercomputing Institute, University of Minnesota, 599 Walter Library, 117 Pleasant St. SE, Minneapolis, MN 55455, USA.}
\address[physicsndsu]{Department of Physics, North Dakota State University, P.O. Box 5566, Fargo, ND 58102, USA.}

\begin{abstract}
A recently introduced particle-based model for fluid dynamics with 
continuous velocities is generalized to model fluids with excluded volume 
effects. This is achieved through the use of biased
stochastic multi-particle collisions which depend on local velocities and 
densities and conserve momentum and kinetic energy.  
The equation of state is derived and criteria for the correct choice 
of collision probabilities are discussed. In particular, it is shown how 
a naive implementation can lead to inconsistent density fluctuations.

\end{abstract}


\end{frontmatter}

\section{Introduction}

Simulation studies of the structure and dynamic properties of complex 
liquids are often complicated by the fact that typical energy scales are   
on the order of the thermal energy and the characteristic structural 
length scales are in the range of nanometers to micrometers. The resulting 
large number of degrees of freedom and disparate length and time scales 
require the use of ``mesoscale'' simulation techniques which  
achieve high computational efficiency by ``averaging out'' irrelevant 
microscopic details while retaining the essential features of the 
microscopic physics on the length scales of interest. 
The fact that the properties of these systems are strongly influenced 
by a delicate interplay between thermal fluctuations, 
hydrodynamic interactions, and possible spatio-temporally varying forces, 
places additional stringent requirements on the simulation protocol. 
A recently introduced particle-based simulation technique 
\cite{male_99}---often called stochastic rotation dynamics (SRD)  
\cite{ihle_01,ihle_04,tuze_03,kiku_03,pool_05},  
multi-particle collision dynamics \cite{ripo_04}, or real-coded 
lattice gas~\cite{inou_02}---is a 
promising algorithm for mesoscale simulations of this type. SRD 
solves the hydrodynamic equations of motion by following the path of 
fluid point-particles in discrete time and continuous space. Efficient 
multi-particle collisions which explicitly conserve momentum and energy 
enable simulations in the microcanonical ensemble, while fully 
incorporating both thermal fluctuations and hydrodynamic interactions.  
Furthermore, its simplicity has made it possible to obtain accurate analytic 
expressions for the transport coefficients which are valid for both large 
and small mean free paths. 

\begin{figure}
\includegraphics[height=2.5in]{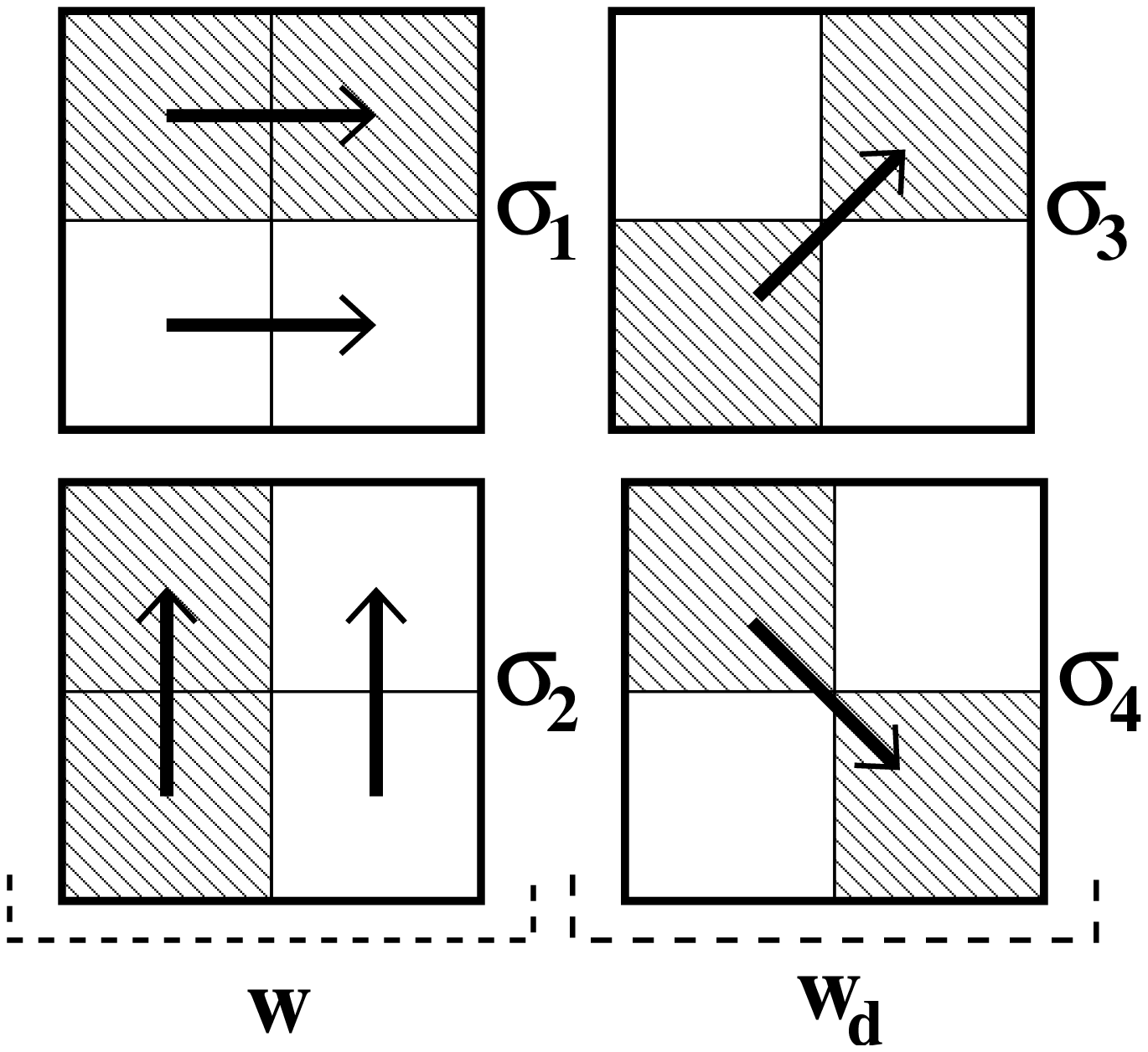}\label{fig_schematic}
\hspace{0.15cm}
\includegraphics[height=2.5in]{pressure_vs_M_and_kt_inset.eps} 
\caption{Collision rules. Four distinct collisions are considered: 
a) horizontally along $\bsigma_1$, b) vertically along $\bsigma_2$, 
c) diagonally and d) off-diagonally along $\bsigma_3$ and $\bsigma_4$.  
$w$ and $w_d$ denote the 
probabilities of choosing collisions a), b) and c), d) respectively.}
\caption{$P_n$ times $\tau$ as a function 
of $M^{1/2}$, measured using the microscopic stress tensor. Symbols show data 
in the range $\tau=0.05, \ldots ,2.00$.
Parameters: $L/a=32$, $k_BT=1.0$. The inset shows the nonideal contribution 
to the pressure as a function of $(k_BT)^{1/2}$.  Parameters: $L/a=32$, $M=10$, 
$\tau=0.40$. In both figures, the solid lines are the theoretical prediction 
given by Eq. (\ref{pressure}).  }
\label{fig_pressure}
\end{figure}

The original SRD algorithm---in which collisions consist of a stochastic 
rotation of the relative velocities of particles in the collision 
cells---describes a fluid with an ideal gas equation of state. 
The fluid is therefore very 
compressible, and the speed of sound, $c_s$, is low. However, this collision 
rule is not unique and other choices of collision rules can lead to nonideal 
behavior. In this paper we consider one such collision rule 
and discuss in detail the conditions which must be fulfilled in order to 
guarantee thermodynamic consistency. There are, however, several subtleties 
associated with collision rules of this type,  
such as phase space contraction and semi-detailed balance in this reduced phase 
space, which are not addressed here. In systems with 
explicit interparticle potentials---such as the hard sphere fluid---this 
behavior can be analyzed in considerable detail. However, to the best of our 
knowledge, there are no non-trivial models with multi-particle interactions 
of the type we consider for which this has been done. The thermodynamic 
consistency of such models is not ensured a priori, and the aim of this paper 
is to provide guidelines for the construction of  consistent models and to 
point out possible pitfalls.

A more realistic modeling of dense gases and liquids can be achieved 
by introducing generalized excluded volume interactions between the fluid 
particles. The resulting algorithm can be thought of as a coarse-grained 
multi-particle collision generalization of a hard sphere fluid, since, 
just as for hard spheres, the kinetic energy is conserved. There is 
no potential energy, so that the internal energy is the same as that of 
an ideal gas. Thermodynamic consistency therefore requires that 
$c_v=T\ ds/dT\vert_\rho = dk_B/2$, where $d$ is the spatial dimension. 
It follows that the nonideal contribution to the entropy density, $s_n$, 
can only depend on the density, $\rho$, so that the free energy density  
$f(T,\rho) = f_{ideal}(T,\rho) + T s_n(\rho)$. The equation of 
state is therefore $P = \rho \partial f/\partial \rho\vert_T - 
f = P_{ideal} + T[\rho \partial s_n(\rho)/\partial\rho-s_n]$, and  
$P-P_{ideal}$ is strictly proportional to the temperature $T$.

\section{Model}

As in the original SRD algorithm, the solvent is modeled by  
$N$ of point-like particles of mass $m$ which move in continuous space with 
a continuous distribution of velocities. The system is coarse-grained 
into $(L/a)^d$ cells of a $d$-dimensional cubic lattice of linear 
dimension $L$ and lattice constant $a$. 
The algorithm consists of individual streaming and collision steps. In 
the free-steaming step, the coordinates, ${\bf r}_i(t)$, of the solvent 
particles at time $t$ are updated according to ${\bf r}_i(t+\tau)=
{\bf r}_i(t) + \tau {\bf v}_i(t)$, where ${\bf v}_i(t)$ is the velocity 
of particle $i$ at time $t$ and $\tau$ is the value of the discretized 
time step. In order to define the collision, we introduce a second grid  
with sides of length $2a$ which (in $d=2$) groups four adjacent cells into one 
``supercell''. For simplicity, we restricted ourselves to two dimensions, 
but the algorithm can be easily extended to three dimensions.

As proposed in Ref. \cite{ihle_01}, a random shift of 
the particle coordinates before the collision step is required to ensure 
Galilean invariance. All particles are therefore shifted by the {\it same}
random vector with components in the interval $[-a,a]$ before the collision 
step. 
Particles are then shifted back by the same amount 
after the collision. To initiate a collision, pairs of cells in every supercell 
are randomly selected. As shown in Fig. 1, three distinct choices are 
possible: a) horizontal ($\bsigma_1$), b) vertical ($\bsigma_2$), and 
c) diagonal collisions ($\bsigma_3$ and $\bsigma_4$). 
In every cell, we define the mean particle velocity,
${\bf u}_n={(1/ M_n)}\,\sum_{i=1}^{M_n}\,{\bf v}_i$, where the sum 
runs over all particles, $M_n$, in the cell with index $n$. 
The projection of the difference of the mean velocities of the selected 
cell-pairs on ${\bsigma}_j$, 
$\Delta u={\bsigma}_j\cdot ({\bf u}_1-{\bf u}_2)$, 
is then used to determine the probability of a collision.
If $\Delta u<0$, no collision will be performed. 
For positive $\Delta u$, a collision will occur with an acceptance 
probability which depends, in principle, on $\Delta u$ and the number 
of particles in the two cells, $M_1$ and $M_2$. Indeed, the choice 
of acceptance probability, $p_A$, determines both the equation of state and 
values of the transport coefficients, and the requirement of thermodynamic 
consistency imposes severe restrictions on the choice of $p_A$. In Ref. 
\cite{ihle_epl} 
it was shown that the choice $p_A(M_1,M_2,\Delta u) = \Theta(\Delta u) 
\tanh(\Lambda)$, with $\Lambda = A\Delta u M_1M_2$, where $\Theta$ is the 
unit step function and A is a (small) constant leads to an equation of 
state of the required form. In this paper we explore the consequences of 
the simpler choice $p_A =  \Theta(\Delta u)$, which is identical to the 
limit $A\rightarrow \infty$ of the model presented in Ref.~\cite{ihle_epl}.

The collisions should conserve 
the total momentum and kinetic energy of the cell-pairs participating in 
the collision, and in analogy to the hard-sphere liquid, they should 
primarily transfer the component of the momentum which is parallel 
to the connecting vector $\bsigma_j$.
The rule we have chosen is to exchange the parallel component of the mean 
velocities of the two cells, which is equivalent to a ``reflection'' of 
the relative velocities~\cite{ihle_epl}.
The perpendicular component remains unchanged.
Because of $x-y$ symmetry, the probabilities for choosing cell pairs in the 
$x-$ and $y-$ directions  are equal, and will be denoted by $w$.
The probability for choosing diagonal pairs  is given by $w_d=1-2w$. $w$ and $w_d$ must be chosen to that the 
hydrodynamic equations are isotropic and do not depend on the orientation 
of the underlying grid.  As shown in Ref. \cite{ihle_epl},  
this can be achieved only if $w_d=1/2$ and $w=1/4$. 

\section{Equation of state and the structure factor}  

\begin{figure}
\label{fig_sound}
\includegraphics[height=2.5in]{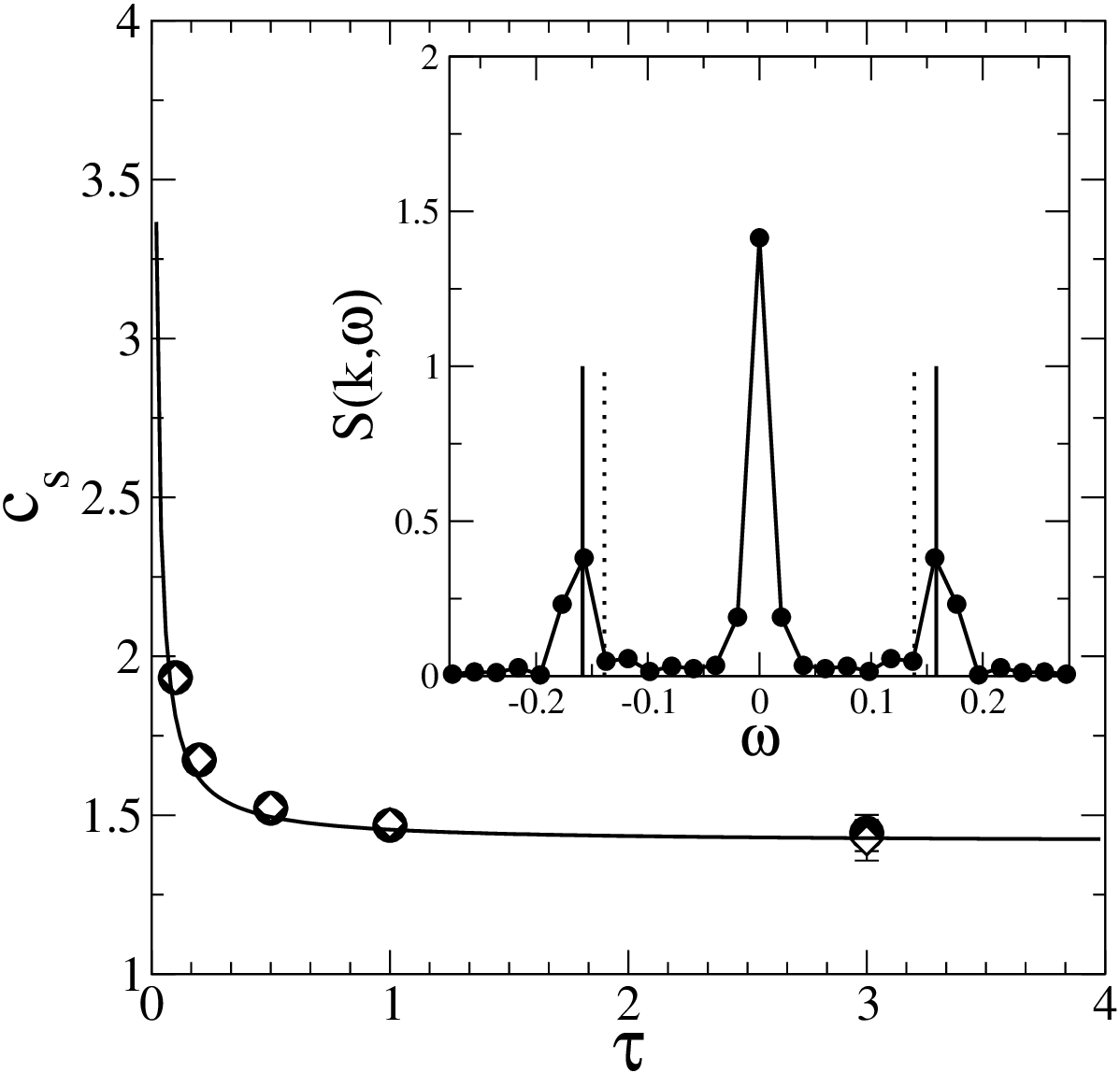}
\hspace{0.2cm}
\includegraphics[height=2.5in]{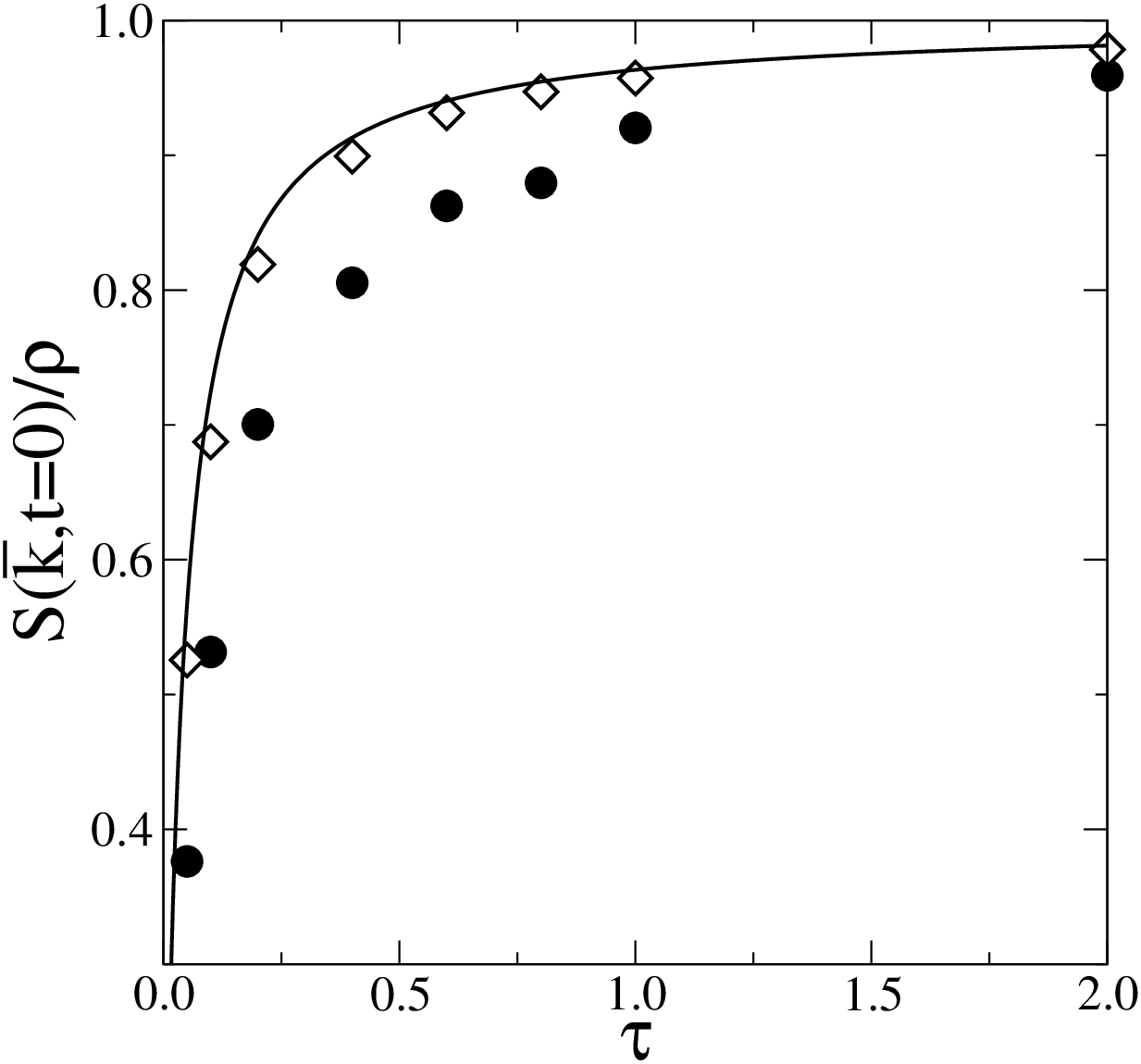} 
\caption{Speed of sound as a function of $\tau$.  (\lbullet) and (\ldiamond) 
are results for $k=(1,0)$ and $(0,1)$, respectively. The solid line is the 
theoretical prediction given by Eq. (\ref{sound}). The inset shows the 
dynamical structure factor as a function of $\omega$ for $k=(2,0)$, $\tau=0.2$. 
Parameters: $L/a=128$, $M=5$, $k_BT=1.0$.}
\caption{$S(\overline{k},t=0)/\rho$ as a function of $\tau$ (\lbullet). 
(\ldiamond) are results obtained by numerically evaluating the derivative of 
the pressure measured using the microscopic stress tensor. 
The solid line is a plot of Eq. (\ref{structure}). 
$\overline{k}$ is the lowest wave vector.}
\label{fig_S}
\end{figure}

The pressure can be calculated using the method described in \cite{ihle_epl}. 
For $p_A= \Theta(\Delta u)$, one finds 
\begin{equation}\label{pres}  
\label{pressure}
P=P_{ideal} + P_n = \rho k_B T+ \left(b/\tau\right) \sqrt{\rho k_B T}
\end{equation}
in the limit of large $M$, where $\rho=M/a^2$ is the particle density and 
$b=(1/4+1/2\sqrt{2})/2\sqrt{\pi}$. The first term in Eq. (\ref{pres}) is 
the ideal gas contribution and the second is the contribution from the 
collisions, the nonideal pressure, $P_n$. 
It can be shown that $P_n\sim \rho^2$ in the limit of small $M$.
Simulation results for $P_n$ obtained by averaging the diagonal part of the 
microscopic stress tensor \cite{ihle_epl} were found to be in good agreement 
with Eq. (\ref{pressure}), see Fig. \ref{fig_pressure}. This equation of state 
is not consistent with the fact that the kinetic energy is conserved, since
$P-P_{ideal}\sim \sqrt{T}$ instead of $T$.
Using Eq. (\ref{pressure}), the adiabatic speed of sound in $d=2$ is  
\begin{equation}
\label{sound}
c_s^2=k_B T\left[1+\frac{b}{2\tau \sqrt{\rho k_BT}}\right]
\left[2+\frac{b}{2\tau \sqrt{\rho k_BT}}\right] \;\;\;.
\end{equation}
We have performed simulations to determine the dynamical structure 
factor, $S(k,\omega)=\langle\rho(k,\omega)\rho(k,\omega)\rangle$. 
Measuring the structure factor can be tricky and details are 
discussed elsewhere. $S(k,\omega)$ 
is plotted as a function of $\omega$ in the inset to Fig. \ref{fig_sound}. 
The position of the finite frequency peaks in  $S(k,\omega)$ gives the 
speed of sound. The solid vertical lines in the figure show the 
theoretically predicted positions of the frequencies for the speed of sound 
given in Eq. (\ref{sound}). The dashed lines show the peak positions for 
an ideal gas. Results for  
the speed of sound for various $k$-values is shown in Fig. \ref{fig_sound}. 
The agreement with Eq. (\ref{sound}) is satisfactory. We have also checked 
that the sound speed is isotropic for the model.
Thermodynamics provides a relation between density fluctuations and the  
derivative of the pressure; i.e. using Eq. (\ref{pressure}), 
\begin{equation}\label{structure} }
S(k,t=0)=\rho k_BT{\left.\frac{\partial\rho}{\partial P}\right|_T =  
\frac{\rho}{1+b/(2\tau\sqrt{\rho k_BT})}\;\;.
\end{equation}

Fig. \ref{fig_S} compares simulation data for $S(k,t=0)/\rho$,  
results for the second expression in Eq. (\ref{structure}) obtained by 
taking the numerical derivative of the measured pressure, and  
the analytical result, the last term in Eq. (\ref{structure}). Data obtained 
by measuring the fluctuations in the density, $S(k,0)$, are clearly not 
consistent with the results based on measurements of the expectation value 
of the diagonal part of the microscopic stress tensor. As already discussed, 
the reason for this is that the equation of state is not consistent with 
the fact that the algorithm conserves kinetic energy. Note that for 
$p_A= \Theta(\Delta u)$, the collision probability does not depend on the 
density, which is clearly unphysical. Moreover, the presence of a 
$\sqrt{\rho}$ term in Eq.(\ref{pressure}) is strange since the first term 
in a typical virial expansion would be proportional to $\rho^2$.
The consequences of choosing collision rules which violate thermodynamic 
consistency are therefore quite dramatic and easy to detect. However, as 
shown in Ref. \cite{ihle_epl}, for the correct choice 
of collision probabilities, the algorithm can be made thermodynamically 
consistent.

Support from the National Science Foundation under Grant No. DMR-0513393
and ND EPSCoR through NSF grant EPS-0132289 are gratefully acknowledged.
We thank A.J. Wagner for numerous discussions.

\end{document}